\documentclass[aps,prd,twocolumn,superscriptaddress]{revtex4}
\usepackage{epsfig,epsf}
\usepackage{amsmath}
\usepackage{amsthm}
\usepackage{amsfonts}
\usepackage{amssymb}
\usepackage{dsfont}
\usepackage{multirow}
\usepackage{appendix}
\usepackage{slashed}
\usepackage[active]{srcltx}
\usepackage{psfrag}

\begin{document}

\title{Mass and residue of $ \Lambda(1405)$ as  hybrid and excited ordinary baryon }
\date{\today}
\author{K.~Azizi}
\affiliation{Department of Physics, Do\v{g}u\c{s} University, Acibadem-Kadik\"{o}y, 34722
Istanbul, Turkey}
\affiliation{School of Physics, Institute for Research in Fundamental Sciences (IPM), P. O. Box 19395-5531, Tehran, Iran}
\author{B.~Barsbay}
\affiliation{Department of Physics, Kocaeli University, 41380 Izmit, Turkey}
\author{H.~Sundu}
\affiliation{Department of Physics, Kocaeli University, 41380 Izmit, Turkey}

\begin{abstract}
The nature of the $ \Lambda(1405) $ has been a puzzle for decades, whether it is a standard three-quark baryon, a hybrid baryon or a baryon-meson molecule. More information on the decay channels of this particle and its strong, weak and electromagnetic interactions with other hadrons is needed to clarify its internal organization. The residue of this particle is one of the main inputs in investigation of its decay properties in many approaches. We calculate the mass and residue of $\Lambda(1405)$  state in the context of QCD sum rules considering it as  a hybrid  baryon with three-quark---one gluon content as well  as an excited ordinary baryon with  quantum numbers  $I(J^{P})=0(1/2^{-})$.
 The comparison of the obtained results on the mass with the  average experimental value presented in PDG allows us to interpret this state as 
 a hybrid baryon.
\end{abstract}

\maketitle

\section{Introduction}

It was already suggested that, in addition to the standard particles, there might exist hadrons with different quark-gluon structures, which cannot be included into the ordinary $q\bar{q}$ and $qqq$ schemes.
Due to their unconventional nature these states were included into a group of  hadrons known as exotic particles \cite{Olsen:2014mea}. The discoveries of the exotic hadrons by various collaborations, and collected experimental information on their mass, width  and decay channels  have made investigation of these states one of the central topics in high energy physics. Starting from the first observation of the $X(3872)$ resonance in 2003 by the Belle Collaboration ~\cite{Choi:2003ue}, numerous experimental groups planned to search for  and detect resonances with unusual properties, and
in  fact, measured their mass and width; and determined their quantum numbers. The hadrons with unusual internal structures, i.e. exotic resonances were classified as $ XYZ $ particles \cite{Shen:2016nth}, glueballs \cite{Pimikov:2016pag}, hybrids \cite{Ho:2016owu,Chanowitz:1982qj}, meson molecules \cite{Tuan:1977bg}, tetraquarks \cite{Goerke:2016hxf}, pentaquarks \cite{Chen:2016otp} and dibaryons \cite{Gal:2014zia}.
Like other groups of exotic particles, an identification and classification of hybrid hadrons  and calculation of their spectroscopic parameters are important  for both revealing their inner organization and gaining new information on quantum chromodynamics (QCD).

The existence of hybrid mesons was first  suggested by Jaffe and Johnson in 1976~\cite{Jaffe:1975fd}. The main ingredients of hybrid mesons ($\bar{q}gq$) are a color-octet quark-antiquark pair and an excited gluonic field. A system with these constituents may have all $J^{PC}$ quantum numbers, implying that   one of the fruitful ways to search for hybrids is to study these states  with exotic quantum numbers, which are
 forbidden for the  $q\bar{q}$ states.
The light hybrid mesons were studied  in the framework of  different theoretical methos,
such as the Bag model, flux tube model, lattice QCD  and QCD sum rules
(see for instance \cite{Barnes:1982tx,Barnes:1995hc,Close:1994hc,Hedditch:2005zf,Jin:2002rw} and references therein).
Unfortunately, the predictions for the masses of the hybrids obtained
within these approaches differ considerably from each others. The properties of the
heavy quarkonium hybrids were also calculated using various methods. Thus, relevant explorations
were carried out
in the constituent gluon model, the flux
tube model,  QCD sum
rules, nonrelativistic
QCD and lattice~ (for instance see  Refs. \cite{Horn:1977rq,Govaerts:1986pp,Kleiv:2014fda,Chen:2013eha,Chen:2014fza,Petrov:1998ha,Liu:2012ze} and references therein). 

Hybrid baryons can be defined in two  different ways: as particles containing three valence quarks and a gluon; and  three quarks moving is an excited adiabatic potential.
Despite clear theoretical definitions, the experimental identification of the hybrid
baryons is a more difficult task compared to the hybrid mesons. Since for the baryons, unfortunately
there are  not  $J^{PC}$ exotics, so one must use other features of these particles to determine whether or not they are hybrids (for more information see Ref.\cite{Barnes:2000vn}).

One of the candidates to hybrid baryons is the $\Lambda(1405)$ baryon, which for many decades has attracted interest of physicists. More than forty years ago experiments showed that there was a state with the spin $1/2$
\cite{Engler:1965zz}, which was predicted to be a $\bar{K}-N$ resonance with quantum numbers
$I(J^P)=0(1/2^-)$ \cite{Dalitz:1967fp} using a SU(3) meson-baryon potential. 
The $\Lambda(1405)$ was  experimentally  observed in the low energy exclusive reactions, where, as usual, the kaon and pion beams were used~\cite{Thomas:1973uh,Hemingway:1984pz}. Recently, many  high-statistics data are available by the LEPS,  CLAS and HADES Collaborations~\cite{Niiyama:2008rt,Moriya:2013hwg,Agakishiev:2012xk}. The  spin-parity $I(J^P)=0(1/2^-)$ was then experimentally confirmed for this particle  by the CLAS collaboration~\cite{Moriya:2014kpv} (for  detailed information, see for instance  Ref.\cite{Cho:2017dcy}).

The mass and different decay properties of  $\Lambda(1405)$ were studied using different theoretical methods including lattice QCD \cite{Jido:2009jf, Hyodo:2008ek,Akaishi:2010wt,An:2010wb,Jido:2010ag,MartinezTorres:2010zv,An:2010tv,Takahashi:2010nj,Sekihara:2010uz,Hyodo:2011ur,Sekihara:2011hw,Menadue:2011pd,MartinezTorres:2012yi,Revai:2012fx,Oller:2013zda,Nakamura:2013boa,Sekihara:2013sma,Menadue:2013xqa,Nakamura:2013qda,Dote:2014ema,Ohnishi:2014iba,Hall:2014uca,Sekihara:2014ica,Hall:2014gqa,Nam:2015yoa,Miyahara:2015bya,Miyahara:2015cja,Miyahara:2015uya,He:2015cca,Ohnishi:2015iaq,Miyahara:2015eyq,Fernandez-Ramirez:2015fbq,Hyodo:2015rnm,Molina:2015uqp,Kamiya:2016jqc,Liu:2016wxq,Dong:2016auh,Hall:2016kou,Kim:2017nxg,Meissner1,Meissner2,Meissner3,Meissner4,Meissner5}. Despite a lot of experimental and theoretical studies on the properties of $\Lambda(1405)$ state, unfortunately, there remain important questions about its nature and internal quark organization whether it is 
a standard three-quark baryon, a hybrid baryon or a baryon-meson molecule with one or two-pole structure. Hence,  more experimental and theoretical studies 
are needed to clarify the physical properties of $\Lambda(1405)$ .

In the present study   we are going to calculate the mass and residue of the $\Lambda(1405)$ considering it as a hybrid baryon  with three-quark---one gluon content  as well as an excited ordinary baryon with quantum numbers $I(J^{P})=0(1/2^{-})$ in the framework of QCD  sum rule.   The mass of this state had already been calculated in Ref.~\cite{Kisslinger:2009dr} using the same method. In Ref.~\cite{Kisslinger:2009dr} the     $\Lambda(1405)$ was also considered in two different pictures: as a hybrid and as a mixed hybrid/three-quark strange baryon. The calculations on the mass of this state and comparison of the obtained result with the experimental data allowed the authors to conclude that this state is consistent with being a strange hybrid baryon rather than a mixed state.     The main aim here is to calculate the residue of this state in the considered pictures besides its mass. The residue is one of the main inputs in calculations of many parameters related to the strong, weak and electromagnetic decays of $\Lambda(1405)$ in many theoretical approaches. Investigations of such decay channels may help us better understand the internal organization of  $\Lambda(1405)$ resonance and hopefully solve the puzzle on its nature.

This work is structured in the following way. In sec. \ref{sec:Mass} we derive two-point QCD sum rules for the mass and residue of the $\Lambda(1405)$ baryon by considering it as a hybrid baryon. In sec. III we perform similar computations by treating the $\Lambda(1405)$ as  an excited ordinary three-quark strange baryon in the $\Lambda$ channel. Section \ref{sec:Conc} is reserved for our  concluding remarks.


\section{$ \Lambda(1405)  $ as a Hybrid Baryon}

\label{sec:Mass}

To calculate the mass and residue of the $\Lambda(1405)$ state in the framework of two-point QCD sum rule, we start with the correlation function
\begin{equation}
\Pi_{\Lambda_{H}}(q)=i\int d^{4}xe^{iq\cdot x}\langle 0|\mathcal{T}\lbrace\eta_{\Lambda_{H}}(x)\bar{\eta}_{\Lambda_{H}}(0)\rbrace|0\rangle ,  \label{eq:CorrF1}
\end{equation}
where ${\eta}_{\Lambda_{H}}$ is the interpolating  current for the hybrid
 $\Lambda(1405)$ baryon and $ \mathcal{T} $ indicates the time ordering operator. In the three-quark---one gluon picture, one of the acceptable
interpolating currents  having the quantum numbers  $I(J^{P})=0(1/2^{-})$  is
\begin{eqnarray}
\eta_{\Lambda_{H}}(x)&=&\frac{1}{\sqrt{2}} \epsilon^{abc}\left\{\left[ u^a(x)
C\gamma^{\mu} s^b(x)\right]  \gamma^{\alpha} \left[G_{\mu \alpha} d(x)\right]^c\right.  \nonumber \\
&-&\left. \left[ d^a(x) C\gamma^{\mu} s^b(x)\right] \gamma^{\alpha}\left[G_{\mu \alpha} u(x)\right]^c\right\} ,\label{eq:CDiq}
\end{eqnarray}
where $ a, b, c $ are color indices, $ C $ is the charge conjugation operator, $ u, d, s $ are light quark fields and
\begin{equation}
G^{\mu\nu}=\sum_{A=1}^8 \frac{\lambda^{A}}{2} G_{A}^{\mu\nu},\
\end{equation}
with $\lambda_{A}$ being the  generators of the color SU(3) group. 

The  correlation function $\Pi_{\Lambda_{H}}(q)$ can be calculated in two different ways. From phenomenological or physical side it is obtained in terms of hadronic parameters. From theoretical or QCD side,
it is evaluated in terms of quark's and gluon's degrees of freedom by the help of the operator product expansion (OPE) in deep Euclidean region.  The QCD sum rules for the physical observables such as the mass and residue are obtained equating the coefficients of the same structure in  both representations of the correlation function.  Finally, the continuum subtraction and  Borel transformation are performed in order to suppress the contribution of the higher states and continuum.

First we calculate the correlation
function in terms of the hadronic degrees of freedom. By inserting a complete set of hadronic state into  Eq. (\ref{eq:CorrF1}) and performing integral
over $x$, we get
\begin{equation}
\Pi_{\Lambda_{H}}^{\mathrm{Had}}(q)=\frac{\langle 0|\eta_{\Lambda_{H}}|\Lambda(q)\rangle \langle \Lambda(q)|\bar{\eta}_{\Lambda_{H}}|0\rangle}{m_{\Lambda}^{2}-q^{2}}+\cdots,
\end{equation}
with $m_{\Lambda}$ being the mass of the $\Lambda(1405)$ state. Here the dots indicate
contributions to the correlation function arising from the higher resonances
and continuum states. We define the residue $\lambda_{\Lambda}$ using the
matrix element
\begin{equation}
\langle 0|\eta_{\Lambda_{H}}|\Lambda(q)\rangle =\lambda_{\Lambda}u_{\Lambda_{H}}(q,s).
\label{eq:Res}
\end{equation}
Then performing the summation over spins in accordance with
\begin{eqnarray}
\sum_{s} u_{\Lambda_{H}} (q,s)  \bar{u}_{\Lambda_{H}} (q,s) &= &\!\not\!{q} + m_{\Lambda},
\end{eqnarray}
for the physical side of the correlation function we get
\begin{equation}
\Pi_{\Lambda_{H}}^{\mathrm{Had}}(q)=\frac{\lambda_{\Lambda}^{2}}{m_{\Lambda}^{2}-q^{2}}(\!\not\!{q} + m_{\Lambda})+\cdots.  \label{eq:CorM}
\end{equation}

The Borel transformation with respect to $ q^{2} $ applied to $\Pi_{\Lambda_{H}}^{\mathrm{Had}}(q)$  leads to the final form of the hadronic representation:
\begin{equation}
\mathcal{\widehat B}_{q^{2}}\Pi_{\Lambda_{H}}^{\mathrm{Had}}(q)=\lambda_{\Lambda}^2 e^{-\frac{m_{\Lambda}^{2}}{M^{2}}} \left( \!\not\!{q}+m_{\Lambda} \right)  +\cdots.  \label{eq:CorBor}
\end{equation}

The OPE side of the correlation function is calculated at large space-like region, where $ q^{2}\ll 0 $ in terms of  quark-gluon degrees of freedom. For this end, we substitute the interpolating current given by Eq. (\ref{eq:CDiq}) into Eq. (\ref{eq:CorrF1}), and contract the relevant quark fields. As a result, we get
\begin{eqnarray}
\Pi_{\Lambda_{H}}^{\mathrm{OPE}}(q)&=&\frac{i}{4}\epsilon_{abc}\epsilon_{a'b'c'}\int d^4
 x e^{iqx}\langle 0 |G_{\mu \alpha}(x) G_{\nu \alpha'}(0)|0 \rangle \nonumber \\
&\times& \left\{(\gamma_{\alpha} S^{cc'}_{d}(x)\gamma_{\alpha'})
Tr\left[ \gamma_{\nu}\widetilde{S}^{aa'}_{u}(x)
\gamma_{\mu}S^{bb'}_{s}(x)\right]\right.
\nonumber \\
&+&\left. (\gamma_{\alpha} S^{cc'}_{u}(x)\gamma_{\alpha'})Tr\left[ \gamma_{\nu}\widetilde{S}^{aa'}_{d}(x)
\gamma_{\mu}S^{bb'}_{s}(x)\right]\right.
 \nonumber \\
&-& \left. \gamma_{\alpha} S^{ca'}_{d}(x)\gamma_{\nu}\widetilde{S}^{bb'}_{s}(x)\gamma_{\mu}S^{ac'}_{u}(x)\gamma_{\alpha'}\right. \nonumber \\
&-& \left. \gamma_{\alpha}S^{ca'}_{u}(x)\gamma_{\nu}\widetilde{S}^{bb'}_{s}(x)\gamma_{\mu}
S^{ac'}_{d}(x)\gamma_{\alpha'}  \right\} . \label{eq:CorrF2}
\end{eqnarray}
where $ Tr[ \lambda^{A}\lambda^{B}]=2 \delta^{AB}$ has been used.
In Eq.\ (\ref{eq:CorrF2}) $S_{s,u,d}^{ab}(x)$  are the
light quarks' propagators and we have used  the notation
\begin{equation}
\widetilde{S}_{s,u,d}(x)=CS_{s,u,d}^{T}(x)C.
\end{equation}
We work with the light quark propagator $S_{q}^{ab}(x)$ defined in the form
\begin{eqnarray}
&&S_{q}^{ab}(x)=i\delta _{ab}\frac{\slashed x}{2\pi ^{2}x^{4}}-\delta _{ab}
\frac{m_{q}}{4\pi ^{2}x^{2}}-\delta_{ab}\frac{\langle \bar{q}q\rangle}{12}  \notag \\
&&+i\delta _{ab}\frac{\slashed xm_{q}\langle \bar{q}q\rangle }{48}
-\delta _{ab}\frac{x^{2}}{192}\langle \bar{q}g_{s}\sigma Gq\rangle
+i\delta _{ab}\frac{x^{2}\slashed xm_{q}}{1152}  \notag \\
&&\times \langle \bar{q}g_{s}\sigma Gq\rangle-i\frac{g_sG_{ab}^{\alpha
\beta }}{32\pi ^{2}x^{2}}\left[ \slashed x{\sigma _{\alpha \beta }+\sigma
_{\alpha \beta }}\slashed x\right]  \notag \\
&&-i\delta_{ab}\frac{x^{2}\slashed xg_{s}^{2}\langle \bar{q}q\rangle
^{2}}{7776}-\delta _{ab}\frac{x^{4}\langle \bar{q}q\rangle \langle
g_{s}^{2}G^2\rangle }{27648}+\cdots.
\label{eq:qprop}
\end{eqnarray}
Let us emphasize that in calculations we set the light quark masses $m_u$
and $m_d$ equal to zero, preserving at the same time dependence of the
propagator $S_{s}^{ab}(x)$ on the $m_s$.

We will treat with $ \langle 0 |G_{\mu \alpha}(x)G_{\nu \alpha'}(0)|0 \rangle $ in Eq.\ (\ref{eq:CorrF2}) in two different ways: first we will replace it by the gluon full-propagator in space representation, i.e,
\begin{eqnarray}
\label{12}
&&\langle 0 |G_{\mu \alpha}(x)G_{\nu \alpha'}(0)|0 \rangle =
\frac{1}{2 \pi^2 x^4} [g_{\alpha \alpha'}(g_{\mu \nu}-\frac{4 x_{\mu} x_{\nu}}{x^2}) \nonumber \\
&& +(\alpha, \alpha') \leftrightarrow (\mu, \nu)
 -\alpha \leftrightarrow \mu -\alpha' \leftrightarrow \nu], \notag \\
 &&{}
\end{eqnarray}
and do all calculations. Such calculations are equivalent to the diagrams with valence-gluon as a full propagator. Secondly, we will write it in terms of gluon condensate using
\begin{eqnarray}
\label{aa}
\langle 0 |G_{\mu \alpha}(x)G_{\nu \alpha'}(0)|0 \rangle =
\frac{\langle
g_{s}^{2}G^2\rangle }{96} [g_{\mu \nu} g_{\alpha \alpha'}-g_{\mu \alpha'} g_{\nu\alpha }],
\end{eqnarray}
which represents the diagrams containing the gluon interacting with the QCD vacuum.

The correlation function $\Pi_{\Lambda_{H}}^{\mathrm{OPE}}(q)$ can be decomposed over the Lorentz structures  $\sim$  $\!\not\!{q}$ and $\sim I$. In calculations, we choose the terms   $\sim$  $\!\not\!{q}$.

The chosen invariant amplitude $\Pi^{\mathrm{OPE}}(q^{2})$ can be written
down as the dispersion integral
\begin{equation}
\Pi^{\mathrm{OPE}}(q^{2})=\int_{m_{s}^{2}}^{\infty }\frac{\rho^{
\mathrm{OPE}}(s)}{s-q^{2}}ds+...,
\end{equation}
where $\rho^{\mathrm{OPE}}(s)$ is the two-point spectral density obtained after lengthy calculations on OPE side and taking the imaginary part of the obtained result. The spectral density corresponding to the structure $\!\not\!{q}$ is obtained as
\begin{equation}
\rho^{\mathrm{OPE}}(s)=\rho ^{\mathrm{pert.}}(s)+\sum_{k=3}^{10}\rho_{k}(s),
\label{eq:A1}
\end{equation}
where $ \rho^{\mathrm{pert.}}(s) $ is the perturbative part of the obtained result and  by $\rho_{k}(s)$ we
denote the nonperturbative contributions to $\rho^{\mathrm{OPE}}(s)$.  The perturbative and nonperturbative parts of the spectral density are obtained as:
\begin{widetext}
\begin{eqnarray}
\rho ^{\mathrm{pert.}}(s)&=&\frac{g_{s}^{2}s^{4}}{491520\pi ^{6}},  \nonumber \\
\rho _{\mathrm{3}}(s)&=&-\frac{g_{s}^{2}m_{s}s^{2} \left[ \langle \bar{d}d\rangle- 2 \langle \bar{s}{s}\rangle+\langle \bar{u}u\rangle\right] }{4096\pi^{4}},  \nonumber \\
\rho_{\mathrm{4}}(s)&=&0,  \nonumber \\
\rho_{\mathrm{5}}(s)&=&\frac{g_{s}^{2}m_{0}^{2}m_{s}s \left[ 3\langle \bar{d}d\rangle-4 \langle \bar{s}s\rangle + 3\langle \bar{u}u\rangle\right] }{6144\pi^{4}},  \nonumber \\
\rho_{\mathrm{6}}(s)&=&\frac{g_{s}^{2}s\left[ \langle \bar{d}d\rangle^{2}g_{s}^{2}+27\pi ^{2}\langle \bar{s}s\rangle\langle \bar{u}u\rangle+ 27 \pi^{2}\langle \bar{d}d\rangle(\langle \bar{s}s\rangle+\langle \bar{u}u\rangle)+g_{s}^{2}(\langle \bar{s}s\rangle^{2}+\langle \bar{u}u\rangle^{2})\right] }{10348\pi^{4}}  , \nonumber \\
\rho_{\mathrm{7}}(s)&=&-g_{s}^{2}\langle \alpha _{s}\frac{G^{2}
}{\pi }\rangle \frac{m_{s}\left[ \langle \bar{d}d\rangle +\langle \bar{s}s\rangle +\langle \bar{u}u\rangle\right]  }{6144\pi^{2}}-\frac{1}{256}\langle \alpha _{s}\frac{
G^{2}}{\pi }\rangle m_{s}\left[ \langle \bar{d}d\rangle+\langle \bar{u}u\rangle \right] ,  \nonumber \\
\rho _{\mathrm{8}}(s)&=&-\frac{g_{s}^{2}m_{0}^{2}\left[ \langle \bar{s}s\rangle\langle \bar{u}u\rangle+\langle \bar{d}d\rangle(\langle \bar{s}s\rangle+\langle \bar{u}u\rangle)\right] }{256\pi ^{2}}, \nonumber \\
\rho _{\mathrm{9}}(s)&=&0, \nonumber \\
\rho _{\mathrm{10}}(s)&=&0.
\label{eq:A2}
\end{eqnarray}

\end{widetext}

Now
applying the Borel transformation to $\Pi^{\mathrm{OPE}}(q^{2})$,
equating the obtained expression with the relevant part of the function $
\mathcal{B}_{q^{2}}\Pi_{\Lambda_H }^{\mathrm{Had}}(q)$, and subtracting the
continuum contribution we get the required sum rules. Thus the
mass of the $\Lambda$ state can be evaluated from the sum rule
\begin{equation}
m_{\Lambda}^{2}=\frac{\int_{m_{s}^{2}}^{s_{0}}dss\rho ^{\mathrm{OPE}
}(s)e^{-s/M^{2}}}{\int_{m_{s}^{2}}^{s_{0}}ds\rho^{\mathrm{OPE}} (s)e^{-s/M^{2}}}.
\label{eq:srmass}
\end{equation}
To extract the residue $\lambda_{\Lambda}$ we can employ the sum rule 
\begin{equation}
\lambda_{\Lambda}^{2}e^{-m_{\Lambda}^{2}/M^{2}}=
\int_{m_{s}^{2}}^{s_{0}}ds\rho^{\mathrm{OPE}}(s)e^{-s/M^{2}}.
\label{eq:srcoupling}
\end{equation}

\begin{table}[tbp]
\begin{tabular}{|c|c|}
\hline\hline
Parameters & Values \\ \hline\hline
$m_{s} $ & $96^{+8}_{-4}~\mathrm{MeV} $ \\
$\langle \bar{q}q \rangle $ & $(-0.24\pm 0.01)^3$ $\mathrm{GeV}^3$ \\
$\langle \bar{s}s \rangle $ & $0.8\ \langle \bar{q}q \rangle$ \\
$m_{0}^2 $ & $(0.8\pm0.1)$ $\mathrm{GeV}^2$ \\
$\langle \bar{s}g_s\sigma Gs\rangle$ & $m_{0}^2\langle \bar{s}s \rangle$
\\
$\langle\frac{\alpha_sG^2}{\pi}\rangle $ & $(0.012\pm0.004)$ $~\mathrm{GeV}^4 $ \\
\hline\hline
\end{tabular}
\caption{Input parameters.}
\label{tab:Param}
\end{table}

The QCD sum rules for the mass and residue of the $\Lambda(1405)$
contain various parameters that should be fixed in accordance with the
standard procedures. Thus, for numerical computation of the $m_{\Lambda}$ and $\lambda_{\Lambda}$ we need values of the quark, gluon and mixed condensates as well as the $s$ quark mass.
The values of these parameters can be found in Table \ref{tab:Param}.
 The  QCD sum rules for the physical quantities under consideration additionally depend
on the continuum threshold $s_{0}$ and Borel parameter $M^{2}$. One needs to fix some regions, where physical quantities are practically independent of or demonstrate weak
dependence on these auxiliary parameters according to the standard prescription. To find the working window for the Borel parameter, we
require the convergence of the operator product expansion as well as
adequate suppression of the contributions arising from the higher resonances and
continuum.  As a result we find the interval
\begin{equation}
1.8\ \mathrm{GeV}^{2}\leq M^{2}\leq 3.6\ \mathrm{GeV}^{2},
\end{equation}
for the Borel mass parameter. Our analyses show that in the interval
\begin{equation}
2.1\,\,\mathrm{GeV}^2 \leq s_{0}\leq 2.3 \,\,\mathrm{GeV}^2,
\end{equation}
the results relatively weakly depend on the continuum threshold $s_{0}$. By varying the parameters $M^2$ and $s_{0}$ within the
allowed ranges, as well as taking into account the errors coming
from other input parameters we estimate uncertainties of the whole
calculations. The  mass $m_{\Lambda}$ and residue $\lambda_{\Lambda}$ 
are depicted as  functions of the Borel
and threshold parameters in Figs.~\ref{fig:Massqslash} and \ref{fig:Residueqslash}. From these figures we see that the results demonstrate good stability  with respect to the helping parameters $M^2$ and $s_{0}$ in their working windows.

\begin{widetext}

\begin{figure}[h!]
\begin{center}
\includegraphics[totalheight=6cm,width=8cm]{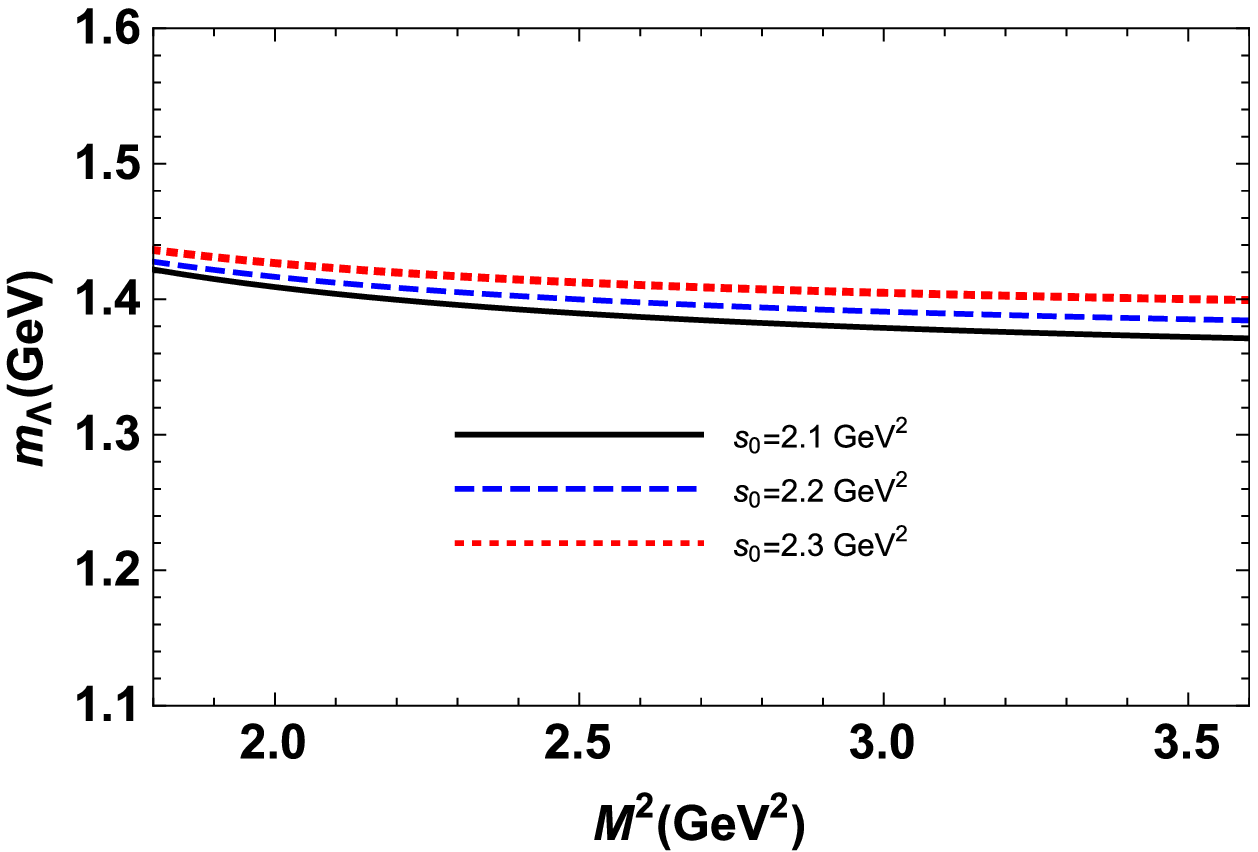}
\,\,
\includegraphics[totalheight=6cm,width=8cm]{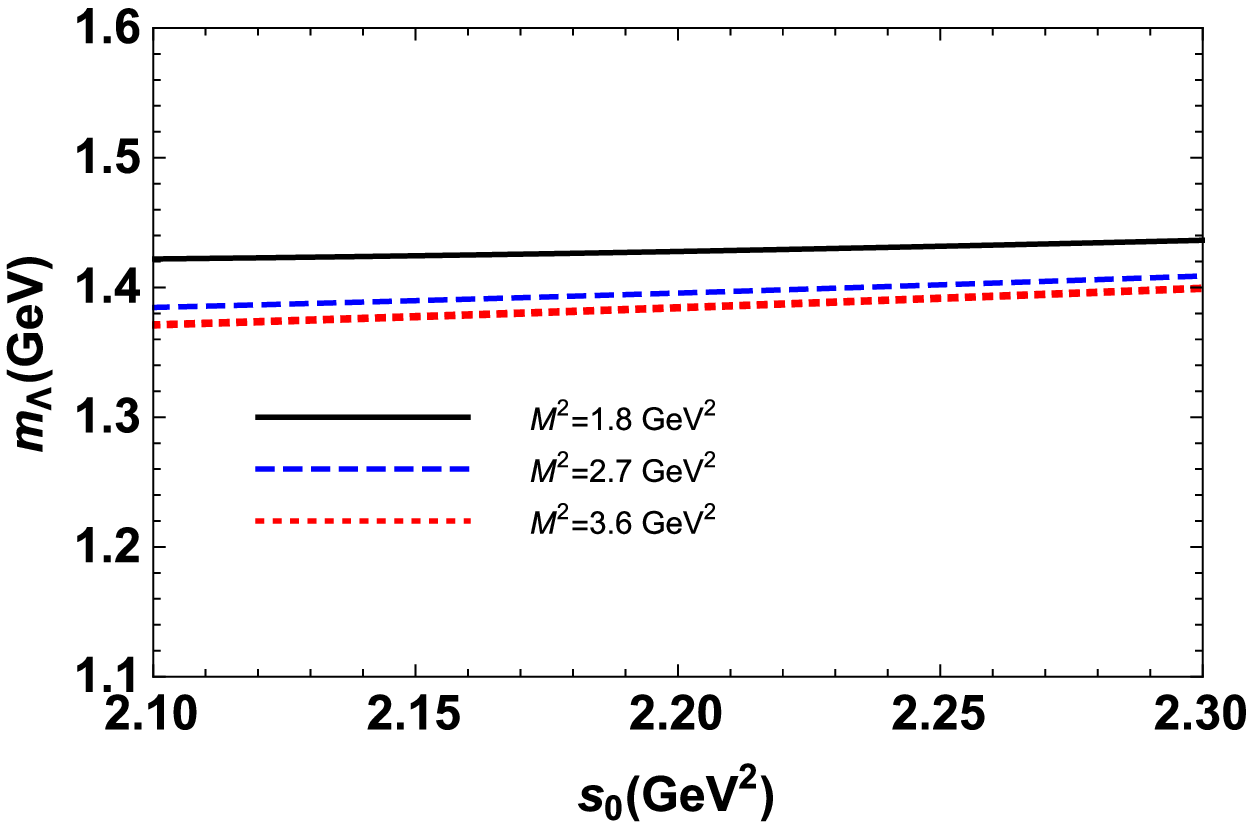}
\end{center}
\caption{ The mass $m_{\Lambda}$ as a function of the Borel parameter $M^2$ at different fixed values of  $s_0$
(left panel), and as a function of the threshold $s_0$ at fixed values of $M^2 $ (right panel).}
\label{fig:Massqslash}
\end{figure}
\begin{figure}[h!]
\begin{center}
\includegraphics[totalheight=6cm,width=8cm]{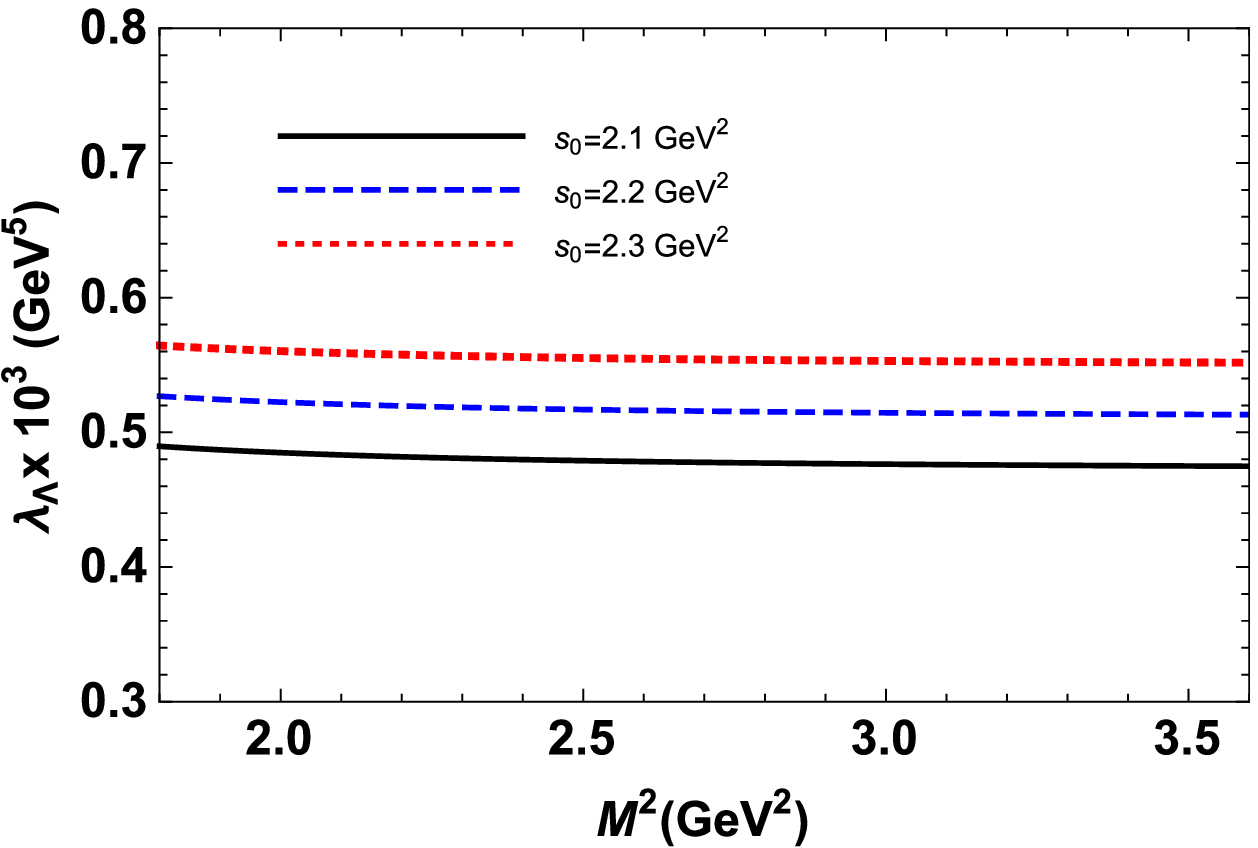}
\,\,
\includegraphics[totalheight=6cm,width=8cm]{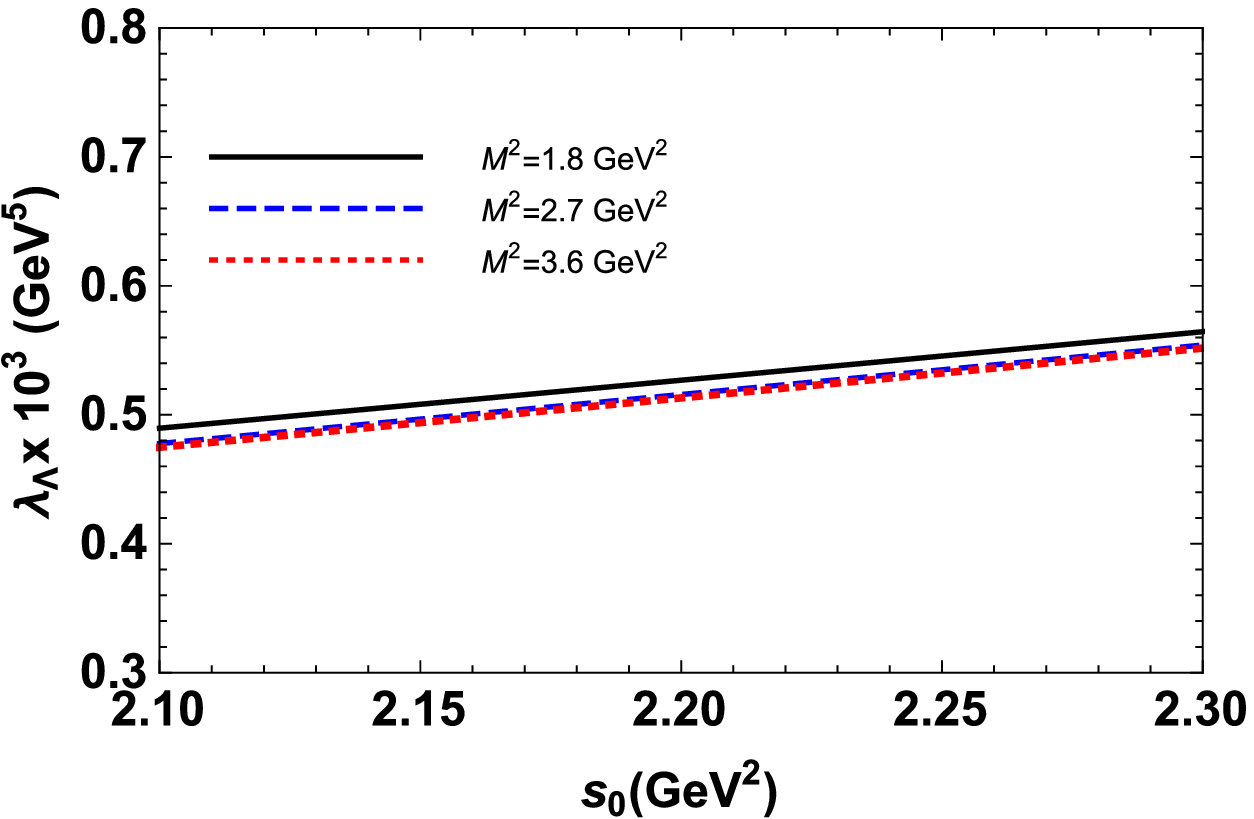}
\end{center}
\caption{ The residue $\lambda_{\Lambda}$ as a function of the Borel parameter $M^2$ at different fixed values of $s_0$
(left panel), and as a function of the threshold $s_0$ at fixed values of $M^2$ (right panel).}
\label{fig:Residueqslash}
\end{figure}

\end{widetext}

Obtained from our analyses, the average values of the mass and residue for $\Lambda(1405)$ are depicted in table  \ref{tab2}.
\begin{table}[tbp]
 \begin{tabular}{|c|c|c|c|}\hline
 &Present Work & \cite{Kisslinger:2009dr}& Experiment \cite{Patrignani}
\\\hline\hline
$m_{\Lambda}$ & $1403^{+33}_{-32}~\mathrm{MeV} $ &$1407~\mathrm{MeV} $ &$(1405.1^{+1.3}_{-1.0})~\mathrm{MeV}$\\
\hline $\lambda_{\Lambda}$& $0.52^{+0.05}_{-0.04}\times 10^{-3}~\mathrm{{GeV}^{5}}$ &$  -$ &$  -$\\
\hline\hline
\end{tabular}
\caption{Values for the mass and residue of
the $\Lambda(1405)$ state.} \label{tab2}
\end{table}
For comparison, we also depict the QCD sum rules prediction on the mass of this state from Ref. \cite{Kisslinger:2009dr} and the average experimental value from PDG \cite{Patrignani} in the same table. From this table we see that our result on the mass is nicely consistent with the prediction of \cite{Kisslinger:2009dr} and the average experimental value. Our prediction on the residue of $\Lambda(1405)$ in the considered picture may be checked via different approaches.

\section{$\Lambda (1405)$ as an excited ordinary three-quark strange baryon}

In this section, we consider the $\Lambda (1405)$ as excited P-wave ordinary three-quark strange baryon with the same quantum numbers as the previous section. As both the ground (with positive parity) and orbitally excited P-wave (with negative parity)  $\Lambda$ baryons  couple to the same current, we will calculate the parameters of both these states and compare with the existing experimental data.
In order to calculate the mass and residue of the positive and the negative
parity spin$-1/2$ $\Lambda $ baryons, we start again with the following two point
correlation function:
\begin{equation}
\Pi (q)=i\int d^{4}xe^{iq\cdot x}\langle 0|\mathcal{T}\Big\{J_{\Lambda }(x)%
\bar{J}_{\Lambda }(0)\Big\}|0\rangle ,  \label{eq:CorrF111}
\end{equation}%
where $J_{\Lambda }(x)$ is the interpolating current for $\Lambda $ state
with spin $J=1/2$.
This interpolating current for ordinary three-quark baryon  has the following form
\begin{eqnarray}
J_{\Lambda } &=&\frac{1}{\sqrt{6}}\varepsilon ^{abc}\Big\{%
2(u^{aT}Cd^{b})\gamma _{5}s^{c}+(u^{aT}Cs^{b})\gamma _{5}d^{c}  \notag \\
&+&(s^{aT}Cd^{b})\gamma _{5}u^{c}+2\beta (u^{aT}C\gamma _{5}d^{b})s^{c}
\notag \\
&+&\beta (u^{aT}C\gamma _{5}s^{b})d^{c}+\beta (s^{aT}C\gamma _{5}d^{b})u^{c}%
\Big\}~,  \label{eq:current}
\end{eqnarray}%
where the
superscript T denotes the transpose operator, and $\beta $ is an arbitrary
parameter with $\beta =-1$ corresponding to the Ioffe current.

To derive the mass sum rules for the $\Lambda $ baryon we calculate this
correlation function again using both the hadronic and OPE languages. By equating
these two representations, one can get the QCD sum rules for the physical
quantities of the  baryon under consideration. The hadronic side of the
correlation function is again obtained by inserting  complete sets of
intermediate states with  both parities. After performing the
four-dimensional  integration over $x$ we get

\begin{eqnarray}
\Pi ^{\mathrm{Phys}}(q) &=&\frac{\langle 0|J|\Lambda (q,s)\rangle \langle
\Lambda (q,s)|\overline{J}|0\rangle }{m^{2}-q^{2}}  \notag \\
&&+\frac{\langle 0|J|\widetilde{\Lambda }(q,\widetilde{s})\rangle \langle
\widetilde{\Lambda }(q,\widetilde{s})|\overline{J}||0\rangle }{\widetilde{m}%
^{2}-q^{2}}  \notag \\
&+&\ldots ,  \label{eq:CF1/2}
\end{eqnarray}%
where $m$, $\widetilde{m}$ and $s$, $\widetilde{s}$ are the masses and spins
of the positive and negative parity  $\Lambda $ baryons, respectively. The
dots denote contributions of higher resonances and continuum states. In Eq.\
(\ref{eq:CF1/2}) the summations over the spins $s$, $\widetilde{s}$ are
implied.

We proceed by introducing the matrix elements
\begin{eqnarray}
\langle 0|J|\Lambda (q,s)\rangle  &=&\lambda u(q,s),  \notag \\
\langle 0|J|\widetilde{\Lambda }(q,\widetilde{s})\rangle  &=&\widetilde{%
\lambda }\gamma _{5}u^{-}(q,\widetilde{s}).  \label{eq:MElem}
\end{eqnarray}%
Here $\lambda $ and $\widetilde{\lambda }$ are the residues of the positive
and negative parity  $\Lambda $ baryons, respectively. Using Eqs.\ (\ref%
{eq:CF1/2}) and (\ref{eq:MElem}) and carrying out summation over the spins
of the  baryons by means of the equality
\begin{eqnarray}
\sum\limits_{s}u(q,s)\overline{u}(q,s) &=&\slashed q+m, \nonumber
\end{eqnarray}%
we obtain%
\begin{equation*}
\Pi ^{\mathrm{Phys}}(q)=\frac{\lambda (\slashed q+m)}{m^{2}-q^{2}}+\frac{%
\widetilde{\lambda }(\slashed q-\widetilde{m})}{\widetilde{m}^{2}-q^{2}}%
+\ldots
\end{equation*}%
The Borel transformation of this expression is:
\begin{eqnarray}
\mathcal{B}\Pi ^{\mathrm{Phys}}(q) &=&\lambda ^{2}e^{-\frac{m^{2}}{M^{2}}}(%
\slashed q+m)  \notag \\
&+&\widetilde{\lambda }e^{-\frac{\widetilde{m}^{2}}{M^{2}}}(\slashed q-%
\widetilde{m}).  \label{eq:Bor1}
\end{eqnarray}

The OPE side of the aforementioned correlation function is again calculated in
terms of the QCD degrees of freedom in deep Euclidean region. After
inserting the explicit form of the interpolating current given by Eq. (\ref%
{eq:current}) into the correlation function in Eq. (\ref{eq:CorrF111}) and
performing contractions via  Wick's theorem, we get the OPE side in terms
of the light quark propagators. By using light quark propagator in the
coordinate space and performing the Fourier and Borel transformations, as
well as applying the continuum subtraction, after  lengthy calculations we
obtain
\begin{equation*}
\mathcal{B}\Pi ^{\mathrm{OPE}}(q)=\mathcal{B}\Pi _{1}^{\mathrm{OPE}}(q)%
\slashed q+\mathcal{B}\Pi _{2}^{\mathrm{OPE}}(q)I.
\end{equation*}%
where, as an example, the $\mathcal{B}\Pi _{1}^{\mathrm{OPE}}(q)$ is given as:
\begin{widetext}

\begin{eqnarray}
\mathcal{B}\Pi^{\mathrm{OPE}}_1(q)&=&\int_{0}^{s_0}e^{-\frac{s}{M^2}}
\frac{1}{128\pi^2}\left\{\frac{s^2(5\beta^2+2\beta+5)}{16\pi^2} +
m_s\left[\langle\bar{s}s\rangle(5\beta^2+2\beta+5)+6(\langle\bar{u}u\rangle
+\langle\bar{d}d\rangle)(1-\beta^2)\right]\right.
\nonumber \\
&+&\left. \frac{\langle g^{2}_{s} GG \rangle
(5\beta^2+2\beta+5)}{16\pi^2}+\frac{m_s(1-\beta^2)}{M^2}
\left[3m_0^2(\langle\bar{d}d\rangle+ \langle\bar{u}u\rangle)
 +\frac{\langle g_s^2
GG\rangle (\langle\bar{d}d\rangle+\langle\bar{u}u\rangle)}{3
M^2}\right]\log\left[\frac{s}{\Lambda^2}\right] \right\}
\nonumber \\
&+&\frac{m_s}{768\pi^2}\left[3m_0^2\left(\langle\bar{u}u\rangle+
\langle\bar{d}d\rangle\right)(1-\beta^2)
 (6\gamma_{E}-13)+8m_0^2\langle\bar{s}s\rangle
(\beta^2+\beta+1)\right]-\frac{1}{24}\left[3\langle\bar{s}s\rangle\left(\langle\bar{u}u\rangle+
\langle\bar{d}d\rangle\right) \right.
\nonumber \\
&\times&\left.
(1-\beta^2)-\langle\bar{d}d\rangle\langle\bar{u}u\rangle
(1-\beta)^2\right]-\frac{m_s}{1536M^2\pi^2}\langle g_s^2
GG\rangle\left[4(\langle\bar{d}d\rangle+\langle\bar{u}u\rangle)(1-\beta^2)
+\langle\bar{s}s\rangle (1+\beta)^2\right]
\nonumber \\
&-&\frac{m_s(1-\beta^2)}{384\pi^2M^2s_0}\langle g_s^2
GG\rangle(\langle\bar{d}d\rangle+\langle\bar{u}u\rangle)
\left(M^2+s_0\log\left[\frac{s_0}{\Lambda^2}\right]\right)e^{-\frac{s_0}{M^2}}
- \frac{1}{96M^2}\left[6m_0^2\langle\bar{s}s\rangle
(\langle\bar{d}d\rangle \right.
\nonumber \\
&+&\left.
\langle\bar{u}u\rangle)(1-\beta^2)+m_0^2\langle\bar{d}d\rangle\langle
\bar{u}u\rangle(1-\beta)^2\right]
+\frac{5m_s(1-\beta^2)}{6144\pi^2M^4}m_0^2\langle g_s^2
GG\rangle(\langle\bar{d}d\rangle+\langle\bar{u}u\rangle).
\end{eqnarray}
Here $\gamma_E \simeq 0.577$ is the Euler constant and $\Lambda$
is a scale parameter. $s_0$ is the continuum threshold, and for
simplicity, we ignored to present the terms containing the $u$ and
$d$ quarks masses and those proportional to $m_0^4$ . Note that we
only ignored to present such terms in the above formula and we
will take into account their contributions in the numerical
calculations.
\end{widetext}

Having calculated both the hadronic and OPE sides of the correlation
function, we match the coefficients of the structures $\slashed q$ and $I$
from these two sides and obtain the following sum rules that will be used to
extract the masses and residues of the ground and first excited states:
\begin{eqnarray}
\lambda^{2}e^{-\frac{m^{2}}{M^{2}}}+\widetilde{\lambda}^{2}e^{-\frac{
\widetilde{m}^{2}}{M^{2}}}&=&\mathcal{B}\Pi _{1}^{\mathrm{OPE}}(q),  \notag
\\
\lambda ^{2}e^{-\frac{m^{2}}{M^{2}}}-\widetilde{\lambda}^{2}e^{-\frac{
\widetilde{m}^{2}}{M^{2}}}&=&\mathcal{B}\Pi _{2}^{\mathrm{OPE}}(q).
\label{eq:MFor1}
\end{eqnarray}

As is seen from Eq. (\ref{eq:MFor1}) in order to obtain the numerical values
of the mass and residue of the orbitally excited  $\Lambda $ baryon, we need
the values of the mass and residue of the ground state. Therefore, we first
calculate the mass and residue of the ground state $\Lambda $ baryon by
choosing an appropriate threshold parameter $s_{0}$ in accordance with the
standard prescriptions.

QCD sum rules contain three auxiliary parameters namely the continuum
threshold $s_0$, Borel parameter $M^2$ and $\beta$ arbitrary parameter. We
find their working windows such that the physical quantities under
consideration be roughly independent of these parameters.

Predictions obtained for the mass and pole residue of the positive and
negative parity ordinary three-quark $\Lambda $ baryons, as well as the working ranges of the
parameters $M^{2}$ and $s_{0}$ are collected in Table \ref{tab:Results1A}.
These results  are obtained by varying the parameter $\beta =\mathrm{tan}%
\theta $ within the limits

\begin{table}[tbp]
\begin{tabular}{|c|c|c|}
\hline
& $\Lambda$ & $\widetilde{\Lambda}$ \\ \hline
$M^2 ~(\mathrm{GeV}^2$) & $1.8-3.6$ & $1.8-3.6$ \\ \hline
$s_0 ~(\mathrm{GeV}^2$) & $1.8-2.0$ & $2.1-2.3$ \\ \hline
$m~(\mathrm{MeV})$ & $1116_{+28}^{-29}$ & $1435^{+32}_{-31}$ \\ \hline
$\lambda \cdot 10^{2} ~(\mathrm{GeV}^3)$ & $1.01^{+0.08}_{-0.07}$ & $%
0.81^{+0.06}_{-0.05}$ \\ \hline
\end{tabular}%
\caption{The sum rule results for the mass and residue of the positive and
negative parity $\Lambda$ baryon with the spin-$1/2$.}
\label{tab:Results1A}
\end{table}
\begin{equation}
-0.9\leq \cos\theta \leq -0.3, \, \, \,0.3\leq \cos\theta \leq 0.9.
\end{equation}

By comparison of the obtained results on the masses with the experimental data presented in PDG we see that the ground state's mass is in  good consistency with the experimental value, however, our prediction on the mass of the $\Lambda (1405)$ as an excited ordinary baryon is considerably high compared to the 
average experimental value presented in PDG.

\section{Concluding Remarks}
\label{sec:Conc}

In this letter we reported the QCD sum rules predictions on the mass and residue of the $\Lambda(1405)$  considering it as a hybrid baryon with 
 three-quark---one gluon as well as an excited ordinary three-quark strange baryon  with quantum numbers $I(J^{P})=0(1/2^{-})$. We  found that the mass of the $\Lambda(1405)$ baryon obtained by considering it as a hybrid baryon  is in a good agreement with the average experimental value. However, we found a mass considerably high compared to the experimental data when we considered it as an excited ordinary three-quark P-wave strange baryon. Hence, our predictions allow us to interpret this state as a hybrid baryon rather than the excited ordinary baryons.  Our predictions for the residue of $\Lambda(1405)$ can be tested via different theoretical approaches and may be used as input information to study the strong, weak and electromagnetic interactions of $\Lambda(1405)$ state with other particles, and determine widths of its various decays. Such calculations, especially investigation of internal charge distribution of the $\Lambda(1405)$ baryon and its multipole moments together with the  comparison of the obtained predictions on its mass and width with  experimental data  will allow us with high confidence level to determine whether the $\Lambda(1405)$ is a hybrid baryon, an excited ordinary three-quark P-wave strange baryon or whether it has other quark-gluon organizations.

\section*{ACKNOWLEDGEMENTS}

K. A. and B. B. thank  T\"{U}B\.{I}TAK for  financial support provided under the grant no: 115F183.


\begin{thebibliography}{99}

\bibitem{Olsen:2014mea} 
  S.~L.~Olsen,
  Hyperfine Interact.\  {\bf 229}, no. 1-3, 7 (2014)
  [arXiv:1403.1254 [hep-ex]].

\bibitem{Choi:2003ue} 
  S.~K.~Choi {\it et al.} [Belle Collaboration],
  Phys.\ Rev.\ Lett.\  {\bf 91}, 262001 (2003)
  [hep-ex/0309032].

\bibitem{Shen:2016nth} 
  C.~Shen,
  AIP Conf.\ Proc.\  {\bf 1735}, 060002 (2016).


\bibitem{Pimikov:2016pag} 
  A.~Pimikov, H.~J.~Lee, N.~Kochelev and P.~Zhang,
  Phys.\ Rev.\ D {\bf 95}, no. 7, 071501 (2017)
  [arXiv:1611.08698 [hep-ph]].



\bibitem{Ho:2016owu} 
  J.~Ho, D.~Harnett and T.~G.~Steele,
  arXiv:1609.06750 [hep-ph].


\bibitem{Chanowitz:1982qj} 
  M.~S.~Chanowitz and S.~R.~Sharpe,
  Nucl.\ Phys.\ B {\bf 222}, 211 (1983)
  Erratum: [Nucl.\ Phys.\ B {\bf 228}, 588 (1983)].


\bibitem{Tuan:1977bg} 
  S.~F.~Tuan,
  Phys.\ Rev.\ D {\bf 15}, 3478 (1977).


\bibitem{Goerke:2016hxf} 
  F.~Goerke, T.~Gutsche, M.~A.~Ivanov, J.~G.~Korner, V.~E.~Lyubovitskij and P.~Santorelli,
  Phys.\ Rev.\ D {\bf 94}, no. 9, 094017 (2016)
  [arXiv:1608.04656 [hep-ph]].


\bibitem{Chen:2016otp} 
  H.~X.~Chen, E.~L.~Cui, W.~Chen, X.~Liu, T.~G.~Steele and S.~L.~Zhu,
  Eur.\ Phys.\ J.\ C {\bf 76}, no. 10, 572 (2016)
  [arXiv:1602.02433 [hep-ph]].



\bibitem{Gal:2014zia} 
  A.~Gal and H.~Garcilazo,
  Nucl.\ Phys.\ A {\bf 928}, 73 (2014)
  [arXiv:1402.3171 [nucl-th]].



\bibitem{Jaffe:1975fd} 
  R.~L.~Jaffe and K.~Johnson,
  Phys.\ Lett.\  {\bf 60B}, 201 (1976).


\bibitem{Barnes:1982tx} 
  T.~Barnes, F.~E.~Close, F.~de Viron and J.~Weyers,
  Nucl.\ Phys.\ B {\bf 224}, 241 (1983).


\bibitem{Barnes:1995hc} 
  T.~Barnes, F.~E.~Close and E.~S.~Swanson,
  Phys.\ Rev.\ D {\bf 52}, 5242 (1995)
  [hep-ph/9501405].

\bibitem{Close:1994hc} 
  F.~E.~Close and P.~R.~Page,
  Nucl.\ Phys.\ B {\bf 443}, 233 (1995)
  [hep-ph/9411301].


\bibitem{Hedditch:2005zf} 
  J.~N.~Hedditch, W.~Kamleh, B.~G.~Lasscock, D.~B.~Leinweber, A.~G.~Williams and J.~M.~Zanotti,
  Phys.\ Rev.\ D {\bf 72}, 114507 (2005)
  [hep-lat/0509106].


\bibitem{Jin:2002rw} 
  H.~Y.~Jin, J.~G.~Korner and T.~G.~Steele,
  Phys.\ Rev.\ D {\bf 67}, 014025 (2003)
  [hep-ph/0211304].


\bibitem{Horn:1977rq} 
  D.~Horn and J.~Mandula,
  Phys.\ Rev.\ D {\bf 17}, 898 (1978).

\bibitem{Govaerts:1986pp} 
  J.~Govaerts, L.~J.~Reinders, P.~Francken, X.~Gonze and J.~Weyers,
  Nucl.\ Phys.\ B {\bf 284}, 674 (1987).


\bibitem{Kleiv:2014fda} 
  R.~Kleiv,
  arXiv:1407.2292 [hep-ph].

\bibitem{Chen:2013eha} 
  W.~Chen, T.~G.~Steele and S.~L.~Zhu,
  J.\ Phys.\ G {\bf 41}, 025003 (2014)
  [arXiv:1306.3486 [hep-ph]].

\bibitem{Chen:2014fza} 
  W.~Chen, T.~G.~Steele and S.~L.~Zhu,
  The Universe {\bf 2}, 13 (2014)
  [arXiv:1403.7457 [hep-ph]].

\bibitem{Petrov:1998ha} 
  A.~A.~Petrov,
  hep-ph/9808347.

\bibitem{Liu:2012ze} 
  L.~Liu {\it et al.} [Hadron Spectrum Collaboration],
  JHEP {\bf 1207}, 126 (2012)
  [arXiv:1204.5425 [hep-ph]].

\bibitem{Barnes:2000vn} 
  T.~Barnes,
  nucl-th/0009011.

\bibitem{Engler:1965zz} 
  A.~Engler, H.~E.~Fisk, R.~w.~Kraemer, C.~M.~Meltzer and J.~B.~Westgard,
  Phys.\ Rev.\ Lett.\  {\bf 15}, 224 (1965).

\bibitem{Dalitz:1967fp} 
  R.~H.~Dalitz, T.~C.~Wong and G.~Rajasekaran,
  Phys.\ Rev.\  {\bf 153}, 1617 (1967).


\bibitem{Thomas:1973uh} 
  D.~W.~Thomas, A.~Engler, H.~E.~Fisk and R.~W.~Kraemer,
  Nucl.\ Phys.\ B {\bf 56}, 15 (1973).

\bibitem{Hemingway:1984pz} 
  R.~J.~Hemingway,
  Nucl.\ Phys.\ B {\bf 253}, 742 (1985).

\bibitem{Niiyama:2008rt} 
  M.~Niiyama {\it et al.},
  Phys.\ Rev.\ C {\bf 78}, 035202 (2008)
  [arXiv:0805.4051 [hep-ex]].


\bibitem{Moriya:2013hwg} 
  K.~Moriya {\it et al.} [CLAS Collaboration],
  Phys.\ Rev.\ C {\bf 88}, 045201 (2013)
  Addendum: [Phys.\ Rev.\ C {\bf 88}, no. 4, 049902 (2013)]
  [arXiv:1305.6776 [nucl-ex]].

\bibitem{Agakishiev:2012xk} 
  G.~Agakishiev {\it et al.} [HADES Collaboration],
  Phys.\ Rev.\ C {\bf 87}, 025201 (2013)
  [arXiv:1208.0205 [nucl-ex]].



\bibitem{Moriya:2014kpv} 
  K.~Moriya {\it et al.} [CLAS Collaboration],
  Phys.\ Rev.\ Lett.\  {\bf 112}, no. 8, 082004 (2014)
  [arXiv:1402.2296 [hep-ex]].


\bibitem{Cho:2017dcy} 
  S.~Cho {\it et al.} [ExHIC Collaboration],
  arXiv:1702.00486 [nucl-th].

\bibitem{Jido:2009jf} 
  D.~Jido, E.~Oset and T.~Sekihara,
  Eur.\ Phys.\ J.\ A {\bf 42}, 257 (2009)
  [arXiv:0904.3410 [nucl-th]].

\bibitem{Hyodo:2008ek} 
  T.~Hyodo, W.~Weise, D.~Jido, L.~Roca and A.~Hosaka,
  Mod.\ Phys.\ Lett.\ A {\bf 23}, 2393 (2008)
  [arXiv:0802.2212 [hep-ph]].


\bibitem{Akaishi:2010wt} 
  Y.~Akaishi, T.~Yamazaki, M.~Obu and M.~Wada,
  Nucl.\ Phys.\ A {\bf 835}, 67 (2010)
  [arXiv:1002.2560 [nucl-th]].


\bibitem{An:2010wb} 
  C.~S.~An, B.~Saghai, S.~G.~Yuan and J.~He,
  Phys.\ Rev.\ C {\bf 81}, 045203 (2010)
  [arXiv:1002.4085 [nucl-th]].


\bibitem{Jido:2010ag} 
  D.~Jido, T.~Sekihara, Y.~Ikeda, T.~Hyodo, Y.~Kanada-En'yo and E.~Oset,
  Nucl.\ Phys.\ A {\bf 835}, 59 (2010)
  [arXiv:1003.4560 [nucl-th]].

\bibitem{MartinezTorres:2010zv} 
  A.~Martinez Torres and D.~Jido,
  Phys.\ Rev.\ C {\bf 82}, 038202 (2010)
  [arXiv:1008.0457 [nucl-th]].

\bibitem{An:2010tv} 
  C.~An and B.~Saghai,
  Int.\ J.\ Mod.\ Phys.\ A {\bf 26}, 619 (2011)
  [arXiv:1008.1177 [nucl-th]].
  
  
\bibitem{Takahashi:2010nj} 
  T.~T.~Takahashi and M.~Oka,
  Prog.\ Theor.\ Phys.\ Suppl.\  {\bf 186}, 172 (2010)
  [arXiv:1009.1790 [hep-lat]].

\bibitem{Sekihara:2010uz} 
  T.~Sekihara, T.~Hyodo and D.~Jido,
  Phys.\ Rev.\ C {\bf 83}, 055202 (2011)
  [arXiv:1012.3232 [nucl-th]].


\bibitem{Hyodo:2011ur} 
  T.~Hyodo and D.~Jido,
  Prog.\ Part.\ Nucl.\ Phys.\  {\bf 67}, 55 (2012)
  [arXiv:1104.4474 [nucl-th]].


\bibitem{Sekihara:2011hw} 
  T.~Sekihara, T.~Hyodo and D.~Jido,
  arXiv:1109.0061 [nucl-th].


\bibitem{Menadue:2011pd} 
  B.~J.~Menadue, W.~Kamleh, D.~B.~Leinweber and M.~S.~Mahbub,
  Phys.\ Rev.\ Lett.\  {\bf 108}, 112001 (2012)
  [arXiv:1109.6716 [hep-lat]].

\bibitem{MartinezTorres:2012yi} 
  A.~Martinez Torres, M.~Bayar, D.~Jido and E.~Oset,
  Phys.\ Rev.\ C {\bf 86}, 055201 (2012)
  [arXiv:1202.4297 [hep-lat]].

\bibitem{Revai:2012fx} 
  J.~Revai,
  Few Body Syst.\  {\bf 54}, 1865 (2013)
  Erratum: [Few Body Syst.\  {\bf 54}, 1877 (2013)]
  [arXiv:1203.1813 [nucl-th]].

\bibitem{Oller:2013zda} 
  J.~A.~Oller,
  Int.\ J.\ Mod.\ Phys.\ Conf.\ Ser.\  {\bf 26}, 1460096 (2014)
  [arXiv:1309.2196 [nucl-th]].

\bibitem{Nakamura:2013boa} 
  S.~X.~Nakamura and D.~Jido,
  PTEP {\bf 2014}, 023D01 (2014)
  [arXiv:1310.5768 [nucl-th]].

\bibitem{Sekihara:2013sma} 
  T.~Sekihara and S.~Kumano,
  Phys.\ Rev.\ C {\bf 89}, no. 2, 025202 (2014)
  [arXiv:1311.4637 [nucl-th]].



\bibitem{Menadue:2013xqa} 
  B.~J.~Menadue, W.~Kamleh, D.~B.~Leinweber, M.~Selim Mahbub and B.~J.~Owen,
  PoS LATTICE {\bf 2013}, 280 (2014)
  [arXiv:1311.5026 [hep-lat]].


\bibitem{Nakamura:2013qda} 
  S.~X.~Nakamura and D.~Jido,
  PoS Hadron {\bf 2013}, 141 (2013)
  [arXiv:1312.6768 [nucl-th]].

\bibitem{Dote:2014ema} 
  A.~Doté and T.~Myo,
  Nucl.\ Phys.\ A {\bf 930}, 86 (2014)
  [arXiv:1406.1540 [nucl-th]].

\bibitem{Ohnishi:2014iba} 
  S.~Ohnishi, Y.~Ikeda, T.~Hyodo, E.~Hiyama and W.~Weise,
  J.\ Phys.\ Conf.\ Ser.\  {\bf 569}, no. 1, 012077 (2014)
  [arXiv:1408.0118 [nucl-th]].


\bibitem{Hall:2014uca} 
  J.~M.~M.~Hall, W.~Kamleh, D.~B.~Leinweber, B.~J.~Menadue, B.~J.~Owen, A.~W.~Thomas and R.~D.~Young,
  Phys.\ Rev.\ Lett.\  {\bf 114}, no. 13, 132002 (2015)
  [arXiv:1411.3402 [hep-lat]].


\bibitem{Sekihara:2014ica} 
  T.~Sekihara and S.~Kumano,
  JPS Conf.\ Proc.\  {\bf 8}, 022006 (2015)
  [arXiv:1411.3414 [hep-ph]].

\bibitem{Hall:2014gqa} 
  J.~M.~M.~Hall, W.~Kamleh, D.~B.~Leinweber, B.~J.~Menadue, B.~J.~Owen, A.~W.~Thomas and R.~D.~Young,
  PoS LATTICE {\bf 2014}, 094 (2014)
  [arXiv:1411.3781 [hep-lat]].


\bibitem{Nam:2015yoa} 
  S.~i.~Nam and H.~K.~Jo,
  arXiv:1503.00419 [hep-ph].

\bibitem{Miyahara:2015bya} 
  K.~Miyahara and T.~Hyodo,
  Phys.\ Rev.\ C {\bf 93}, no. 1, 015201 (2016)
  [arXiv:1506.05724 [nucl-th]].


\bibitem{Miyahara:2015cja} 
  K.~Miyahara, T.~Hyodo and E.~Oset,
  Phys.\ Rev.\ C {\bf 92}, no. 5, 055204 (2015)
  [arXiv:1508.04882 [nucl-th]].

\bibitem{Miyahara:2015uya} 
  K.~Miyahara and T.~Hyodo,
  JPS Conf.\ Proc.\  {\bf 10}, 022011 (2016)
  [arXiv:1508.07707 [nucl-th]].


\bibitem{He:2015cca} 
  J.~He and P.~L.~Lu,
  Int.\ J.\ Mod.\ Phys.\ E {\bf 24}, no. 11, 1550088 (2015)
  [arXiv:1510.00580 [nucl-th]].

\bibitem{Ohnishi:2015iaq} 
  S.~Ohnishi, Y.~Ikeda, T.~Hyodo and W.~Weise,
  Phys.\ Rev.\ C {\bf 93}, no. 2, 025207 (2016)
  [arXiv:1512.00123 [nucl-th]].

\bibitem{Miyahara:2015eyq} 
  K.~Miyahara and T.~Hyodo,
  arXiv:1512.02735 [nucl-th].

\bibitem{Fernandez-Ramirez:2015fbq} 
  C.~Fernandez-Ramirez, I.~V.~Danilkin, V.~Mathieu and A.~P.~Szczepaniak,
  Phys.\ Rev.\ D {\bf 93}, no. 7, 074015 (2016)
  [arXiv:1512.03136 [hep-ph]].

\bibitem{Hyodo:2015rnm} 
  T.~Hyodo,
  AIP Conf.\ Proc.\  {\bf 1735}, 020012 (2016)
  [arXiv:1512.04708 [hep-ph]].



\bibitem{Molina:2015uqp} 
  R.~Molina and M.~Döring,
  Phys.\ Rev.\ D {\bf 94}, no. 5, 056010 (2016)
  Addendum: [Phys.\ Rev.\ D {\bf 94}, no. 7, 079901 (2016)]
  [arXiv:1512.05831 [hep-lat]].


\bibitem{Kamiya:2016jqc} 
  Y.~Kamiya, K.~Miyahara, S.~Ohnishi, Y.~Ikeda, T.~Hyodo, E.~Oset and W.~Weise,
  Nucl.\ Phys.\ A {\bf 954}, 41 (2016)
  [arXiv:1602.08852 [hep-ph]].


\bibitem{Liu:2016wxq} 
  Z.~W.~Liu, J.~M.~M.~Hall, D.~B.~Leinweber, A.~W.~Thomas and J.~J.~Wu,
  Phys.\ Rev.\ D {\bf 95}, no. 1, 014506 (2017)
  [arXiv:1607.05856 [nucl-th]].

\bibitem{Dong:2016auh} 
  F.~Y.~Dong, B.~X.~Sun and J.~L.~Pang,
  arXiv:1609.08354 [nucl-th].

  
\bibitem{Hall:2016kou} 
  J.~M.~M.~Hall, W.~Kamleh, D.~B.~Leinweber, B.~J.~Menadue, B.~J.~Owen and A.~W.~Thomas,
  Phys.\ Rev.\ D {\bf 95}, no. 5, 054510 (2017)
  [arXiv:1612.07477 [hep-lat]].
  
 

\bibitem{Kim:2017nxg} 
  S.~H.~Kim, S.~i.~Nam, D.~Jido and H.~C.~Kim,
  arXiv:1702.08645 [hep-ph].

\bibitem{Meissner1}
J.~A.~Oller and U.~G.~Meissner,
Phys.\ Lett.\ B {\bf 500} (2001) 263
[hep-ph/0011146].


\bibitem{Meissner2}
D.~Jido, J.~A.~Oller, E.~Oset, A.~Ramos and U.~G.~Meissner,
Nucl.\ Phys.\ A {\bf 725} (2003) 181
[nucl-th/0303062].


\bibitem{Meissner3}
Z.~H.~Guo and J.~A.~Oller,
Phys.\ Rev.\ C {\bf 87} (2013) no.3,  035202
[arXiv:1210.3485 [hep-ph]].

\bibitem{Meissner4}
M.~Mai and U.~G.~Meissner,
Nucl.\ Phys.\ A {\bf 900} (2013) 51
[arXiv:1202.2030 [nucl-th]].


\bibitem{Meissner5}
M.~Mai and U.~G.~Meissner,
Eur.\ Phys.\ J.\ A {\bf 51} (2015) no.3,  30
[arXiv:1411.7884 [hep-ph]].



\bibitem{Kisslinger:2009dr} 
  L.~S.~Kisslinger and E.~M.~Henley,
  Eur.\ Phys.\ J.\ A {\bf 47}, 8 (2011)
  [arXiv:0911.1179 [hep-ph]].

\bibitem {Patrignani}
C. Patrignani et al.(Particle Data Group), Chin. Phys. C, 40, 100001 (2016).
%

\end{thebibliography}
\end{document}